# Influence of Coriolis force on propagation and reflection of long waves

A.V. Astashenok[1], A.A. Zaitsev

Atlantic Branch P.P. Shirshov Institute, Institute of Oceanology, Russian Academy of Science

Prospekt Mira 1, 236001 Kaliningrad, Russia

**Abstract**

Influence of Coriolis force on propagation and reflection of the stimulated long waves generated pulsing and harmonic concentrated sources is studied. Geostrophic stream are found, is received asymptotics for transition mode. It is established that in case of presence of rectilinear boundary as a result of reflection Kelvin wave propagating along a boundary is formed.

## Introduction

In geophysical hydrodynamics a few precisely solved theoretical problems which solution is expressed by compact analytical formulas due to what there is possible a simple physical analysis of the process. Problems concern to them about influence of Coriolis force on propagation and reflection of the long waves considered in [1 – 3].

Apparently, this theme has not received development though, in our opinion, it deserves more a attention. In [1] cases of action of a pulsing and harmonic source in the boundless sea are considered. It is established that in both cases geostrophic stream is formed. In the second case there are steady-stated waves with frequency of the source propagating from a source on perpetuity. Besides there is transition mode. But the used technique is difficult enough. The question on simplification of a technique of the solution is important as used in [2] technique give the erroneous results; in particular there in a spectrum of the reflected wave Kelvin wave is not outlined.

We repeatedly consider these problems with the purpose to simplify and add the solution given in [1 – 3].

## Basic assumptions

---

[1] e-mail: artemii@albertina.ru

Let's assume that flat reservoir of depth $h$ filled with an ideal incompressible fluid rotates with constant angular velocity $\omega_0$ around of a vertical axis. The axis $z$ of the cartesian coordinate system coincides with an axis of reservoir rotation. The equations of the stimulated linear long waves in such fluid look like [4]

$$u_t - fv + g\eta_x = 0, v_t + fu + g\eta_y = 0, \eta_t + h(u_x + v_y) = J, \qquad (1)$$

where $u(x,y,t), v(x,y,t)$ – components of velocity of fluid motion along axes $x$ and $y$; $\eta(x,y,t)$ – displacement of free surface of fluid; $f = 2\omega_0$ – Coriolis parameter (inertial frequency); $J(x,y,t)$ – distribution of the sources generating long waves. Let up to the moment of time $t=0$ the fluid is in a rest, i. e. $(\eta, u, v)$, $t < 0$, then the sources are actuated. For the further statement it is convenient to define representing function $S(x, y, t)$ through which $\eta$, $u$, $v$ are expressed as follows

$$\eta = S_{tt} + f^2 S, u = -g(S_{tx} + fS_y), v = -g(S_{ty} - fS_x) \qquad (2)$$

It is easy to check up that in this case the first and second equations in (1) are satisfied automatically and last equation is reduced to non-uniform two-dimensional Klein –Gordon equation:

$$S_{tt} + f^2 S - c^2 (S_{xx} + S_{yy}) = J_i, \qquad (3)$$

where it is designated $J_i = H(t) * J = \int_0^t J(t_0) dt_0$ and $c^2 = gh$; here $H(t)$ – Heaviside function. The solution of the Eq. (3) satisfying a condition of causality is given by convolution:

$$S = J_i * E, \qquad (4)$$

in which $E$ – the fundamental solution of two-dimensional Klein – Gordon equation satisfying a condition of causality, that is

$$E_{tt} + f^2 E - c^2 (E_{xx} + E_{yy}) = \delta(x,y,t), \quad E = \frac{H(ct-r)\cos(f\sqrt{c^2t^2 - r^2}/c)}{2\pi c\sqrt{c^2t^2 - r^2}}. \qquad (5)$$

Let's consider also potential vorticity $\sigma(x,y,t)$ which is expressed as follows:

$$\sigma = h(u_y - v_x) + f\eta. \qquad (6)$$

Using Eqs. (1) it is possible to obtain the equation for potential vorticity:

$$\sigma_t = fJ. \qquad (7)$$

From Eq. (7) follows, that in case of absence of sources value of potential vorticity in each point of fluid does not vary in due course. However with the help of potential vorticity it is possible to outline geostrofic stream for various distributions of sources and the limited reservoir with boundaries of an any configuration directly, not solving initial Eqs. (1).

### Geostrophic stream

In geostrophic stream fields are stationary therefore in case of absence of sources the Eqs. (1) becomes:

$$fv + g\eta_x = 0,\ fu + g\eta_y = 0,\ u_x + v_y = 0, \qquad (8)$$

Then, Eqs. (2) which give expression $\eta$, $u$, $v$ through representing function become simpler:

$$\eta = f^2 S,\ u = -gfS_y,\ v = gfS_x. \qquad (9)$$

For potential vorticity we have

$$\sigma = f(f^2 S - c^2(S_{xx} + S_{yy})). \qquad (10)$$

Now it is possible to conclude: let any time the sources creating spatial distribution of yield $J_i$ which does not depend on time operate; then a bit later after the ending of their action formating of geostrophic stream is finished, fields in which are defined through representing function under Eqs. (9), and representing function will be the solution of Helmholtz equation

$$f^2 S - c^2 (S_{xx} + S_{yy}) = J_i, \tag{11}$$

The solution of this equation is given by convolution of function $-J_i$ with the fundamental solution of Helmholtz equation

$$E_G = -\frac{1}{2\pi c^2} K_0(fr/c) \tag{12}$$

where $K_0(r)$ - Macdonald function of the zero order.

This feature of geostrophic stream does enough procedure of determination of geostrophic stream generated by sources in limited reservoir with boudaries of an any configuration. Here it is possible to take advantage of solving of the appropriate problems on reflection and diffraction of monochromatic acoustic and electromagnetic waves as the last also are reduced to the solving of Helmholtz equation. Simplifies procedure that here is not required to use a Sommerfeld condition of radiation. Besides from equalities (9) follows that representing function of geostrophic stream be converted in zero on reservoir boundary. For example, using a method of reflections and the formula (12) we receive that Green functions in cases of half-space $y > 0$, quarter of a plane $x > 0$, $y > 0$, sector $0 < y < \text{tg}(\pi/n)x$ will be the following:

$$G_G(x, y, x_0, y_0) = -\frac{1}{2\pi c^2} K_0\left(f\sqrt{(x-x_0)^2 + (y-y_0)^2}/c\right) +$$

$$+ \frac{1}{2\pi c^2} K_0\left(f\sqrt{(x-x_0)^2 + (y+y_0)^2}/c\right); \tag{13}$$

$$G_G(x, y, x_0, y_0) = -\frac{1}{2\pi c^2} K_0\left(f\sqrt{(x-x_0)^2 + (y-y_0)^2}/c\right) + \frac{1}{2\pi c^2} K_0\left(f\sqrt{(x-x_0)^2 + (y+y_0)^2}/c\right)$$

$$+ \frac{1}{2\pi c^2} K_0\left(f\sqrt{(x+x_0)^2 + (y-y_0)^2}/c\right) - \frac{1}{2\pi c^2} K_0\left(f\sqrt{(x+x_0)^2 + (y+y_0)^2}/c\right); \tag{14}$$

$$G_G(x, y, x_0, y_0) = -\frac{1}{2\pi c^2} \sum_{m=0}^{n-1} K_0\left(f\sqrt{r^2 + r_0^2 - 2rr_0 \cos(\varphi - \varphi_0 - 2\pi m/n)}/c\right) +$$

$$+ \frac{1}{2\pi c^2} \sum_{m=0}^{n-1} K_0\left(f\sqrt{r^2 + r_0^2 - 2rr_0 \cos(\varphi + \varphi_0 - 2\pi m/n)}/c\right). \tag{15}$$

**Pulsing concentrated source in the boundless sea**

In case of the pulsing concentrated source which acting in the beginning of coordinates and having power $Q$, the Eq. (3) becomes:

$$S_{tt} + f^2 S - c^2(S_{xx} + S_{yy}) = QH(t)\delta(x,y). \tag{16}$$

For the solving according to (4) shall to have:

$$S = \frac{Q}{2\pi c} \int_0^t \frac{H(c\tau - r)\cos(f\sqrt{c^2\tau^2 - r^2}/c)d\tau}{\sqrt{c^2\tau^2 - r^2}}. \tag{17}$$

The elementary changing of variables is obtained (17) to a kind:

$$S = \frac{QH(ct-r)}{2\pi c^2} \int_0^{\sqrt{c^2t^2-r^2}/r} \frac{\cos(fr\tau/c)d\tau}{\sqrt{\tau^2+1}}. \tag{18}$$

The Eq. (18) shows, that perturbation comes at the moment $t=r/c$ in a point of observation on distance $r$. At $t > r/c$ it can be presented as

$$S = S_g + S_t,$$

where

$$S_g = \frac{Q}{2\pi c^2} K_0(fr/c), \tag{19}$$

$$S_t = -\frac{Q}{2\pi c^2} \int_{\sqrt{c^2t^2-r^2}/r}^{+\infty} \frac{\cos(fr\tau/c)d\tau}{\sqrt{\tau^2+1}}. \tag{20}$$

The formula (19) describes of geostrophic stream, formula (20) – transition mode. Integration in parts gives the asymptotic of transient at $\sqrt{c^2t^2 - r^2}/r \gg 1$:

$$S_t \sim \frac{Q\sin(f\sqrt{c^2t^2 - r^2}/c)}{2\pi f c^2 t}. \tag{21}$$

It is inertial waves (their frequency is close to Coriolis parameter) [5] which amplitude decreases proportionally time.

The solution (18) represents Green function of a problem (1). For sources of more complex existential configuration the solution obtains by convolution of function $J$ with function (18).

**Harmonic concentrated source in the boundless sea**

In case of the harmonic concentrated source with frequency $\omega$ which acting in the beginning of coordinates $J=QH(t)\exp(i\omega t)\delta(x,y)$. Representing function will obtain by convolution of function $H(t)\exp(i\omega t)$ with function (18) and will be the following:

$$S = \frac{QH(ct-r)}{2\pi c^2}\exp(i\omega t)\int_{r/c}^{t}\int_{0}^{\sqrt{c^2 t_0^2 - r^2}/r}\frac{\cos(fr\tau/c)d\tau}{\sqrt{\tau^2+1}}\exp(-i\omega t_0)dt_0.$$

Here external integration is carried out elementary. In result we obtain:

$$S = \frac{QH(ct-r)}{2\pi i\omega c^2}\int_{0}^{\sqrt{c^2 t^2 - r^2}/r}\frac{(\exp(i\omega(t - r\sqrt{\tau^2+1}/c)) - 1)\cos(fr\tau/c)d\tau}{\sqrt{\tau^2+1}}. \tag{22}$$

At $t > r/c$ equality (22) can be presented as
$$S = S_g + S_s + S_t,$$
where

$$S_g = \frac{iQ}{2\pi c^2 \omega}K_0(fr/c), \tag{23}$$

$$S_s = \frac{Q}{2\pi i\omega c^2}\exp(i\omega t)\int_{0}^{+\infty}\frac{\exp(-i\omega r\sqrt{\tau^2+1}/c)\cos(fr\tau/c)d\tau}{\sqrt{\tau^2+1}}, \tag{24}$$

$$S_t = \frac{iQ}{2\pi\omega c^2}\int_{\sqrt{c^2 t^2 - r^2}/r}^{+\infty}\frac{(\exp(i\omega(t - r\sqrt{\tau^2+1}/c)) - 1)\cos(fr\tau/c)d\tau}{\sqrt{\tau^2+1}}. \tag{25}$$

The formula (23) describes of geostrophic stream, the formula (24) – the steady-stated waves, the formula (25) – transition mode. The integral in (24) is calculated with the help of known formulas for integrated representation of modified Bessel functions [6]:

$$\int_0^{+\infty} \frac{\exp(-i\omega\sqrt{\tau^2+1})\cos(f\tau)d\tau}{\sqrt{\tau^2+1}} = -i\frac{\pi}{2}H_0^{(2)}(\sqrt{\omega^2-f^2}),\ f<\omega,$$

$$\int_0^{+\infty} \frac{\exp(-i\omega\sqrt{\tau^2+1})\cos(f\tau)d\tau}{\sqrt{\tau^2+1}} = K_0(\sqrt{f^2-\omega^2}),\ \omega<f,$$

where $H_0^{(2)}(x)$ – Hankel function of the second kind of the zero order.
Result will be the following formulas:

$$S_s = \frac{Q}{4\pi\omega c^2}\exp(i\omega t)H_0^{(2)}(\sqrt{\omega^2-f^2}\,r/c),\ f<\omega,$$

$$S_s = \frac{Q}{2\pi i\omega c^2}\exp(i\omega t)K_0(\sqrt{f^2-\omega^2}\,r/c),\ \omega<f. \tag{26}$$

Thus the steady-stated displacement represents a progressive or standing wave with the frequency equal to source frequency. It is possible to find asymptotic of transition mode at $\sqrt{c^2t^2-r^2}/r \gg 1$:

$$S_t \sim -\frac{Q}{2\pi f\omega c\sqrt{c^2t^2-r^2}}\cos(f\sqrt{c^2t^2-r^2}/c) + \frac{Q}{2\pi if\omega c^2 t}\sin(f\sqrt{c^2t^2-r^2}/c). \tag{27}$$

The formula (27) shows that transition mode represents running inertial waves with amplitude slowly dumped in due course, in inverse proportion $t$. It is obvious that the real parts of formulas (23) – (25) define an displacement caused by a source with yield $Q\cos(\omega t)$, imaginary – with yield $Q\sin(\omega t)$. Using Eqs. (2) from (18) and (22) it is possible to find an displacement of a free surface and components of fluid velocity.

Separately we shall consider a case of concurrence of sourse frequency with inertial (resonance). In this case equations (26) for the steady-stated mode are inapplicable, since at $\omega = f$ Hankel and Macdonald function of the zero order have singularities. However we are interested with physical quantities, instead of representing function. For source function in this case we have $J = QH(t)\exp(ift)\delta(x,y)$. Let's work on both parts of the Eq. (3) by operator $\partial_{tt} + f^2$. We obtain non-uniform Klein – Gordon equation for an displacement of a free surface:

$$\eta_{tt} + f^2\eta - c^2(\eta_{xx}+\eta_{yy}) = QH(t)*(J_{0tt}+f^2J_0)\delta(x,y), \tag{28}$$

where $J_0 = H(t)\exp(ift)$. Differentiating $J_0$ in sense of the generalized functions, we obtain that $J_{0tt} + f^2 J_0 = \delta'(t) + if\delta(t)$, and hence

$$\eta = Q\left( E + i\frac{f}{2\pi c^2}\int_0^{\sqrt{c^2 t^2 - r^2}/r} \frac{\cos(fr\tau/c)d\tau}{\sqrt{\tau^2+1}} \right). \tag{29}$$

**Pulsing concentrated source in the sea with rectilinear coast**

Let's consider the problem on the stimulated long waves which are generated by the pulsing concentrated source with power $Q$ in the sea with a rectilinear coast $y=0$. We count that waves are propagated in the sea medium which fills half-space $y>0$. The source is in a point $(x_0, y_0)$, $y_0 > 0$. Then representing function satisfies to the Eq. (16) and also a boundary condition

$$S_{ty} - fS_x = 0, \, y = 0; \tag{30}$$

the condition (30) follows from a condition $v=0$ on a coast, and last from representations (2). For the solving of a problem we use a method of reflections. We search function $S$ as

$$S = S^0(x-x_0, y-y_0, t) - S^0(x-x_0, y+y_0, t) + S^r(x-x_0, y+y_0, t), \tag{31}$$

where $S^0(x,y,t)$ – the solution of a problem on action of the pulsing concentrated source which gives the formula (18), that is

$$S^0(x,y,t) = \frac{QH(ct-r)}{2\pi c^2}\int_0^{\sqrt{c^2 t^2 - r^2}/r} \frac{\cos(fr\tau/c)d\tau}{\sqrt{\tau^2+1}}, \, S_t^0 = QE.$$

Function $S^r(x,y,t)$ satisfies to uniform Klein – Gordon equation

$$S_{tt}^r + f^2 S^r - c^2(S_{xx}^r + S_{yy}^r) = 0, \, y > 0. \tag{32}$$

After substitution of representation (31) in a condition (30) it is obtained one more equation for function $S^r(x,y,t)$:

$$S^r_{ty} - fS^{r0}_x = 2QE_y, \quad y > 0. \tag{33}$$

Here are taken into account validity of condition (30) for all positions of a source and evenness of function $S^0(x,y,t)$. We act on the equation (33) operator $c^2(\partial_{ty} + f\partial_x)$ and exclude $S^r_{yy}$ with the help of Klein – Gordon equation:

$$\eta^r_{tt} - c^2\eta^r_{xx} = 2Q(E_{tt} + f^2 E - c^2 E_{xx})_t + 2Qfc^2 E_{xy}, \tag{34}$$

Further let's assume that $\eta^r = 2QE_t + 2Qf\eta_1$, where $\eta_1$ satisfies to the equation

$$\eta_{1tt} - c^2\eta_{1xx} = fE_t + c^2 E_{xy}. \tag{35}$$

For its solving we shall search $\eta_1$ as $\eta_1 = \eta_2 - \eta_3$ where the component $\eta_2$ satisfies to the equation

$$c^2 \eta_{2yy} - f^2 \eta_2 = fE_t + c^2 E_{xy}. \tag{36}$$

The equation for $\eta_3$ is easy for obtaining, using the Eqs. (35), (36) and a Eq. (5) for the fundamental solution of Klein – Gordon equation:

$$(c^2\partial_{yy} - f^2)(\partial_{tt} - c^2\partial_{xx})\eta_3 = (f\partial_t + c^2\partial_{xy})\delta(x,y,t) \tag{37}$$

Let's solve the equation (37) in area y > 0. Taking into account that the fundamental solution of the equation $(c^2\partial_{yy} - f^2)F(y) = \delta(y)$ at positive $y$ is $-\exp(-fy/c)/2fc$ and also that $\partial_{tt} - c^2\partial_{xx} = (\partial_t - c\partial_x)(\partial_t + c\partial_x)$ it is possible to simplify Eq. (37):

$$\eta_{3t} + c\eta_{3x} = -\frac{\delta(x,t)}{2c}\exp(-fy/c). \tag{38}$$

The solution of the given equation also gives Kelvin wave propagating along a coast:

$$\eta_3 = -H(t)\frac{\exp(-fy/c)}{2c}\delta(ct-x). \qquad (39)$$

Let's proceed to procedure of the solving of the equation (36). The solution can be presented as $\eta_2 = (f\partial_t + c^2\partial_{xy})\tilde{\eta}_2$, where $\tilde{\eta}_2$ satisfies to the equation

$$c^2\tilde{\eta}_{2yy} - f^2\tilde{\eta}_2 = E. \qquad (40)$$

The restricted solution of Eq. (40) may be written down in the following form:

$$\tilde{\eta}_2 = -\frac{1}{4\pi fc^2}(\int_{-\infty}^{y} dy_0 \frac{H(ct-\sqrt{x^2+y_0^2})}{\sqrt{c^2t^2-(x^2+y_0^2)}}\cos(f\sqrt{c^2t^2-x^2-y_0^2}/c)\exp(f(y_0-y)/c)$$
$$+\int_{y}^{\infty} dy_0 \frac{H(ct-\sqrt{x^2+y_0^2})}{\sqrt{c^2t^2-(x^2+y_0^2)}}\cos(f\sqrt{c^2t^2-x^2-y_0^2}/c)\exp(-f(y_0-y)/c)). \qquad (41)$$

Two cases are possible:

1) $y > \sqrt{c^2t^2-x^2}$, i.e. $r>ct$. In this case integral (41) it is possible to present as

$$\tilde{\eta}_2 = \frac{-H(ct-|x|)}{4\pi fc^2}\int_{-\sqrt{c^2t^2-x^2}}^{\sqrt{c^2t^2-x^2}} dy_0 \frac{\cos(f\sqrt{c^2t^2-(x^2+y_0^2)}/c)}{\sqrt{c^2t^2-(x^2+y_0^2)}}\exp(-f(y_0-y)/c).$$

This integral can be calculated precisely. For this purpose we make changing of a integration variable $y_0 = \sqrt{c^2t^2-x^2}\cos\varphi$ and we take advantage of standard methods of the theory of functions of a complex variable. In result we obtain

$$\tilde{\eta}_2 = -H(ct-|x|)\frac{\exp(-fy/c)}{4fc^2}, r \geq ct.$$

In this case $\eta_2 = -\frac{\exp(-fy/c)}{4c^2}(\partial_t - c\partial_x)H(ct-|x|)$. Differentiation in sense of the generalized functions in a result gives for $\eta_2$ at $r \geq ct$ in accuracy the same expression, as well as for Kelvin wave (39). Thus at $r \geq ct$ in the spectrum of the reflected wave Kelvin wave does not contain.

This result is physically expected, in time *t* perturbation from a source may not be propagated to distance the greater than *ct*.

2) $0 < y < \sqrt{c^2t^2 - x^2}$, i.e. *r<ct*. Let's transform integral (41), using the same changing of a integration variable as in the previous case. We have:

$$\tilde{\eta}_2 = -\frac{H(ct-|x|)}{4\pi fc^2}(\int_{\arccos(y/\sqrt{c^2t^2-x^2})}^{\pi} d\varphi \cos(f\sqrt{c^2t^2-x^2}\sin\varphi/c)\exp(f\sqrt{c^2t^2-x^2}\cos\varphi/c)\exp(-fy/c)$$

$$+ \int_0^{\arccos(y/\sqrt{c^2t^2-x^2})} d\varphi \cos(f\sqrt{c^2t^2-x^2}\sin\varphi/c)\exp(-f\sqrt{c^2t^2-x^2}\cos\varphi/c)\exp(fy/c)), r < ct.$$

Let's write out final result for $\eta_1$:

$$\eta_1 = H(ct-r)\delta(ct-x)\exp(-fy/c)/2c -$$
$$-\frac{1}{4\pi fc^2}(f\partial_t + c^2\partial_{xy})(H(ct-r)\int_{\arccos(y/\sqrt{c^2t^2-x^2})}^{\pi} d\varphi \cos(f\sqrt{c^2t^2-x^2}\sin\varphi/c)e^{f\sqrt{c^2t^2-x^2}\cos\varphi/c}e^{-fy/c} +$$
$$+H(ct-r)\int_0^{\arccos(y/\sqrt{c^2t^2-x^2})} d\varphi \cos(f\sqrt{c^2t^2-x^2}\sin\varphi/c)e^{-f\sqrt{c^2t^2-x^2}\cos\varphi/c}e^{fy/c}).$$

The result for an displacement of a free surface can be written down in other form. For this purpose we take into account that Fourier transformation on *y* of the fundamental solution of Klein – Gordon equation is [6]

$$\overline{E}(x,k,t) = \frac{H(ct-|x|)}{2c}J_0(\sqrt{k^2+f^2/c^2}\sqrt{c^2t^2-x^2}), \qquad (42)$$

where $J_0(x)$ – Bessel function of the zero order. In a result (41) it is possible to rewritten as follows:

$$\tilde{\eta}_2 = \frac{H(ct-|x|)}{4\pi c}\int_{-\infty}^{\infty} dk \frac{J_0(\sqrt{k^2+f^2/c^2}\sqrt{c^2t^2-x^2})}{f^2+c^2k^2}\exp(-iky). \qquad (41a)$$

At last we write out result for $\eta$:

$$\eta = \frac{Q}{2\pi}\{\frac{\delta(ct-r_1)}{\sqrt{c^2t^2-r_1^2}} - H(ct-r_1)(\frac{ct\cos(f\sqrt{c^2t^2-r_1^2}/c)}{(c^2t^2-r_1^2)^{3/2}} + \frac{ft\sin(f\sqrt{c^2t^2-r_1^2}/c)}{c^2t^2-r_1^2} +$$

$$+\frac{f^2}{c^2}\int_0^{\sqrt{c^2t^2-r_1^2}/r_1}\frac{\cos(fr_1\tau/c)d\tau}{\sqrt{\tau^2+1}}) - \frac{\delta(ct-r_1)}{\sqrt{c^2t^2-r_1^2}} + H(ct-r_2)(\frac{ct\cos(f\sqrt{c^2t^2-r_2^2}/c)}{(c^2t^2-r_1^2)^{3/2}} +$$

$$+\frac{ft\sin(f\sqrt{c^2t^2-r_2^2}/c)}{c^2t^2-r_2^2} - \frac{f^2}{c^2}\int_0^{\sqrt{c^2t^2-r_2^2}/r_2}\frac{\cos(fr_2\tau/c)d\tau}{\sqrt{\tau^2+1}}) + \quad (43)$$

$$+\frac{f(f\partial_t+c^2\partial_{xy})H(ct-|x-x_0|)}{c}\int_{-\infty}^{\infty}dk\frac{J_0(\sqrt{k^2+f^2/c^2}\sqrt{c^2t^2-(x-x_0)^2})}{f^2+c^2k^2}e^{-ik(y+y_0)}\} +$$

$$+H(t)\frac{Qf}{c}\exp(-f(y+y_0)/c))\delta(ct-x+x_0),$$

where is designated $r_1 = \sqrt{(x-x_0)^2+(y-y_0)^2}$ и $r_2 = \sqrt{(x-x_0)^2+(y+y_0)^2}$.

First four terms in the solution (43) correspond to a direct wave, the others – to the waves reflected from boundary.

The solution (43) represents the fundamental solution of a problem (1) with a condition that normal velocity should vanish on rectilinear coastal boundary. For more complex source function $J$ the solution obtains by convolution. For the sources distributed in space, the displacement of a free surface will not have infinite breaks on forward front as these features will be eliminated as a result of integration on coordinates.

For example, in case of the harmonic concentrated source solution for displacement will obtain of convolution of function $H(t)\exp(i\omega t)$ with function (43). There is no necessity completely to give the solution of this problem. Let's outline only Kelvin wave $\eta_k$:

$$\eta_k = H(t)\frac{Qf}{c^2}\exp(-f(y+y_0)/c)\exp(i\omega(t-x/c)). \quad (44)$$

Obviously, formula (44) describes the harmonic wave propagating along a coast in a positive direction with a source frequency.

**Pulsing concentrated source in the channel**

The result obtained in the previous section suggests that it is possible to expect occurrence of Kelvin waves in case of boundaries of more complex configuration. With the purpose of confirmation of the given hypothesis we shall consider propagation of the stimulated waves in the reclinear channel of width $d$. Let the concentrated source with power $Q$ is in a point $(0, y_0)$, $0 < y_0 < d$. For the solving of Eqs. (1) we shall take advantage of Laplace transform on time with parameter $p$ and Fourier transform on $x$ with parameter $k$. The Eqs. (1) with boundary conditions become:

$$\begin{aligned}
& p\bar{u} - f\bar{v} - ikg\bar{\eta} = 0, \\
& p\bar{v} + f\bar{u} + g\bar{\eta}_y = 0, \\
& p\bar{\eta} + h(-ik\bar{u} + \bar{v}_y) = Q\delta(y - y_0), \\
& \bar{v} = 0, \ y = 0, d.
\end{aligned} \quad (45)$$

Here $\bar{u}, \bar{v}, \bar{\eta}$ – images of components of velocity and displacement of a free surface accordingly. We exclude $\bar{u}$ and $\bar{v}_y$ from third equation with the help of first two. As a result of simple calculations we obtain the equation for $\bar{\eta}$:

$$\bar{\eta}_{yy} - \frac{R^2}{c^2}\bar{\eta} = -Q\frac{p^2 + f^2}{c^2 p}\delta(y - y_0), \quad (46)$$

where $R = \sqrt{p^2 + f^2 + c^2 k^2}$. The solution of the equation (46) should be continuous in a point $y_0$ and have the saltus of a derivative equal $-Q\frac{p^2 + f^2}{c^2 p}$. From these conditions we find that

$$\bar{\eta} = \frac{Q}{2cp(p^2 + c^2 k^2)} G, \quad (47)$$

where it is designated

$$G = ((p^2 R^2 + f^2 c^2 k^2)\cosh(R(|y - y_0| - d)/c) + (p^2 R^2 - f^2 c^2 k^2)\cosh(R(y + y_0 - d)/c) - \\ - 2ifckpR\sinh(R(y + y_0 - d)/c))/R\sinh(Rd/c).$$

Further we shall take advantage of known expansion in a series [6]:

$$\frac{\cosh(az)}{z\sinh(z)} = \frac{1}{z^2} + 2\sum_{n=1}^{\infty}\frac{(-1)^n \cos(\pi n a)}{z^2 + \pi^2 n^2}. \quad (48)$$

If differentiate both parts of expansion (48) on $a$ it is possible to obtain one more series:

$$\frac{\sinh(az)}{\sinh(z)} = -2\pi \sum_{n=1}^{\infty} \frac{(-1)^n n \sin(\pi n a)}{z^2 + \pi^2 n^2}. \qquad (49)$$

Using expansions (48), (49) it is possible to expand function $G$ in a series. In a result for an image of an displacement of a free surface we obtain that $\bar{\eta} = \sum_{n=0}^{\infty} \bar{\eta}_n$, where

$$\bar{\eta}_n = \frac{QG_n(p)}{2pd^2(p^2 + c^2k^2)(p^2 + \omega_n^2)},$$
$$G_n(p) = \varepsilon_n (-1)^n (d(p^2 R^2 + f^2 c^2 k^2) \cos(\pi n(|y - y_0| - d)/d) + \qquad (50)$$
$$+ d(p^2 R^2 - f^2 c^2 k^2) \cos(\pi n(y + y_0 - d)/d) + 2\pi i f c^2 k p n \sin(\pi n(y + y_0 - d)/d)).$$

Here $\omega_n^2 = f^2 + c^2(k^2 + \pi^2 n^2 / d^2)$, $\varepsilon_0 = 1$, $\varepsilon_m = 2$, $m \neq 0$. It is possible to expand $\bar{\eta}_n$ to partial fractions:

$$\bar{\eta}_n = \frac{A_n}{p} + \frac{B_n}{p - ick} + \frac{B_n^+}{p + ick} + \frac{C_n}{p - i\omega_n} + \frac{C_n^+}{p + i\omega_n} \qquad (51)$$

After inverse Laplace transform on $p$ and inverse Fourier transform on $k$ it is easy to interpret terms in the expansion (51). The first term corresponds geostrophic stream, the second and the third – to Kelvin waves, the fourth and the fifth – to waves with frequency $\omega_n$.

Let's receive expressions for geostrophic stream and Kelvin waves. For geostrophic displacement $\eta_g$ we have

$$\eta_g = \frac{1}{2\pi} \int_{-\infty}^{\infty} \sum_{n=0}^{\infty} A_n dk \exp(-ikx), \qquad (52)$$

where factors $A_n$ are defined by

$$A_n = \frac{QG_n(0)}{2d^2 c^2 k^2 \omega_n^2} = \frac{Qf^2 d}{c^2}(-1)^n \frac{\cos(\pi n(|y - y_0| - d)/d) - \cos(\pi n(y + y_0 - d)/d)}{\pi^2 n^2 + k^2 d^2 + f^2 d^2 / c^2}.$$

Summation in the equation (52) is easy for executing using expansion (48). After integration we obtain result for $\eta_g$

$$\eta_g = \frac{Qf^2}{c^2 d}\sum_{n=1}^{\infty}\left(\frac{f^2}{c^2}+\frac{\pi^2 n^2}{d^2}\right)^{-1/2}\exp\left(-|x|\left(\frac{f^2}{c^2}+\frac{\pi^2 n^2}{d^2}\right)^{1/2}\right)\sin\left(\frac{\pi n y_0}{d}\right)\sin\left(\frac{\pi n y}{d}\right). \quad (53)$$

From (53) follows that geostrophic displacement convert in zero on boundaries of the channel and quickly dampes along an axis $x$ with increase of distance from a source.

For Kelvin wave $\eta_k^+$ propagating in a positive direction we have

$$\eta_k^+ = \frac{1}{2\pi}\int_{-\infty}^{\infty}\sum_{n=0}^{\infty} B_n \exp(ikct - ikx)dk, \quad (54)$$

where factors $B_n$ are defined by

$$B_n = \frac{QG_n(ick)}{2(ick)^2 d^2(\omega_n^2 - c^2 k^2)} = \frac{Qf^2 \varepsilon_n (-1)^n \cos(\pi n(y+y_0-d)/d)}{2dc^2(f^2/c^2 + \pi^2 n^2/d^2)} +$$
$$+ \frac{Q\pi n f \varepsilon_n (-1)^n \sin(\pi n(y+y_0-d)/d)}{2cd^2(f^2/c^2 + \pi^2 n^2/d^2)}.$$

Summation in (54) is execute with the help expansions (48), (49). After integration we obtain result for $\eta_k^+$:

$$\eta_k^+ = H(t)\frac{Qfe^{-f(y+y_0-d)/c}}{2c\sinh(fd/c)}\delta(ct-x). \quad (55)$$

Formula (55) describes a wave propagating along boundary $y=0$ and damping with distance from it. Similar representation can be obtained for Kelvin wave $\eta_k^-$ propagating in a negative direction:

$$\eta_k^- = H(t)\frac{Qfe^{f(y+y_0-d)/c}}{2c\sinh(fd/c)}\delta(ct+x). \quad (56)$$

Formula (56) describes a wave propagating along boundary $y=d$ and damping with distance from it.

## Conclusion

In the present paper influence of Coriolis force on propagation and reflection of long waves is studied. Initiating of function (2) for Eqs. (1) has allowed to obtain for representing function non-uniform Klein – Gordon equation (3). For the sources working during a finite time interval, the simple way of a finding of geostrophic stream is offered.

Consideration of process of reflection of a wave does not apply for completeness. Obviously, the stated technique allows to study reflection of a wave from boundaries of more complex configuration than a rectilinear coast or channel. It is possible to expect in this case occurrence in a spectrum of the reflected wave of Kelvin waves propagating along boundaries and exponential damping with distance from its.

## References

1. B.I. Sebekin. *Dokl. AN USSR*, Vol. 187, № 1, p. 53. (Russian).
2. S.S. Voit, B.I. Sebekin. *Dokl. AN USSR*, Vol. 191, № 5, p. 1007 (Russian).
3. L.G. Chambers. *Jour. Fluid Mech.*, Vol. 22, № 2.
4. L.N. Sretenski. *Theory of wave motion of fluid*. Moscow – Leningrad, 1936 (Russian).
5. M.J. Lighthill. *Waves in fluids*. Cambridge University Press, 1978.
6. M. Abramowitz, I.A. Stegun. *Handbook of Mathematical Functions, National Bureau of Standards*. Applied Math. Series #55, Dover Publications, 1965.